# Direct Observation of Early-stage Quantum Dot Growth Mechanisms with High-temperature Ab Initio Molecular Dynamics

Lisi Xie[1], Qing Zhao[1,2], Klavs F. Jensen[1] and Heather J. Kulik[1],*

[1]*Department of Chemical Engineering, Massachusetts Institute of Technology, Cambridge, MA 02139*
[2]*Department of Mechanical Engineering, Massachusetts Institute of Technology, Cambridge, MA 02139*

ABSTRACT: Colloidal quantum dots (QDs) exhibit highly desirable size- and shape-dependent properties for applications from electronic devices to imaging. Indium phosphide QDs have emerged as a primary candidate to replace the more toxic CdSe QDs, but production of InP QDs with the desired properties lags behind other QD materials due to a poor understanding of how to tune the growth process. Using high-temperature ab initio molecular dynamics (AIMD) simulations, we report the first direct observation of the early stage intermediates and subsequent formation of an InP cluster from separated indium and phosphorus precursors. In our simulations, indium agglomeration precedes formation of In-P bonds. We observe a predominantly intercomplex pathway in which In-P bonds form between one set of precursor copies while the carboxylate ligand of a second indium precursor in the agglomerated indium abstracts a ligand from the phosphorus precursor. This process produces an indium-rich cluster with structural properties comparable to those in bulk zinc-blende InP crystals. Minimum energy pathway characterization of the AIMD-sampled reaction events confirms these observations and identifies that In-carboxylate dissociation energetics solely determine the barrier along the In-P bond formation pathway, which is lower for intercomplex (13 kcal/mol) than intracomplex (21 kcal/mol) mechanisms. The phosphorus precursor chemistry, on the other hand, controls the thermodynamics of the reaction. Our observations of the differing roles of precursors in controlling QD formation strongly suggests that the challenges thus far encountered in InP QD synthesis optimization may be attributed to an overlooked need for a cooperative tuning strategy that simultaneously addresses the chemistry of both indium and phosphorus precursors.

# 1. Introduction

Colloidal semiconductor nanocrystals (quantum dots, QDs) exhibit unique size- and shape-dependent properties that have been harnessed for applications ranging from optoelectronic and photovoltaic devices[1-4] to bio-imaging reagents[5-6]. First-principles simulations have provided important insights into the unusual structure-property relationships of QDs.[7-10] Cadmium selenide (CdSe)-based QDs are the most widely investigated materials and have been commercialized in consumer electronic products.[11] However, the high toxicity of cadmium[12] has inspired research into replacement materials that have similar electronic and optical properties, and indium phosphide (InP) has been identified as the most promising candidate[13-16]. Despite significant experimental effort to identify the mechanism[17-24] and tune the growth process[25-31] of InP QDs, current synthesis approaches have not yet obtained optimal InP QD size distribution, quantum yield and stability comparable to CdSe-based QDs[32]. Difficulties in optimizing InP QD synthesis have motivated further investigation into the QD growth process, but spectroscopic techniques that can be used to characterize InP clusters/QDs have not been able to identify the difficult-to-isolate early-stage intermediates. The most common experimental recipe[22, 24] for InP QD synthesis involves the use of long-chain carboxylate indium precursors and various phosphorus precursors (Figure 1). First-principles simulations that are intrinsically limited to short time- and length-scales are ideally suited to shed light on the short-lived intermediates in the early-stage growth of InP QDs.

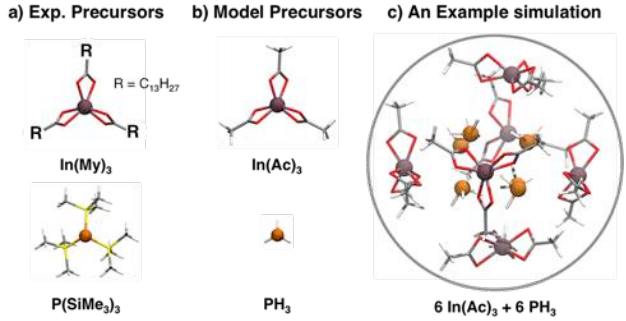

**Figure 1.** (a) Experimental precursors (indium myristate (In(My)$_3$) and tris(trimethylsilyl) phosphine (P(SiMe$_3$)$_3$)), (b) model precursor analogues used in simulations (indium acetate (In(Ac)$_3$), phosphine (PH$_3$)), and (c) snapshot from example ab initio molecular dynamics simulation with spherical boundary depicted. Indium (brown) and phosphorus (orange) atoms are shown as spheres with their annotated partial charges, while other atoms (oxygen in red, silicon in yellow, carbon in gray, and hydrogen in white) are shown in stick representation.

First-principles simulations have been widely employed to study primarily optoelectronic properties of the more established II-VI and IV-VI cadmium[33-39] or lead chalcogenide[10, 37, 40-42] QDs, although there have been a few computational studies of III-V InP QDs[43-46]. In addition to structure-property investigations, binding energies to crystalline models of possible QD facets have been used to propose growth mechanisms of QDs.[33-34, 36, 41, 47] For example, relative ligand binding strengths have been used to understand anisotropic growth in CdSe QDs[33-34], the effect of surface oxidation or hydroxylation on the growth of CdSe[36] and PbS[41] QDs, and properties of facets of indium oxide nanoparticles[47]. While InP QDs have not been the subject of as much computational study, other InP materials such as nanowires[48], nanotubes[49], hetero-structures[50] and bulk surfaces[51-53] have been investigated with first-principles techniques. Mechanisms of InP nanowire growth on InP crystals have been proposed[54] based on relative binding energies for various In and P adatom configurations. Experimentally, more is known about the growth mechanism of metal nanoparticles (e.g. gold), leading to considerable computational effort[55-61] of formation reaction energetics and ligand exchange mechanisms[58-61].

The challenges in InP QD synthesis motivate the use of first-principles simulations to investigate early-stage growth intermediates, but, to the best of our knowledge, no such computational studies have been carried out thus far. The lack of experimental information about the nature of InP QD growth intermediates necessitates a sampling approach to first discover possible pathways rather than directly evaluating energetics for predefined pathways. Ab initio molecular dynamics (AIMD) has been widely employed as a discovery tool where mechanisms were not already known including, for example, the growth of carbon nanotube or graphene on metal nanoparticles or surfaces[62-64], water splitting on InP/GaP surfaces[51-53], CO oxidation on ceria-supported gold nanoclusters[65] and water oxidation by cobalt nanoparticles[66]. These AIMD simulations are constrained to short timescales and small system sizes in order to avoid prohibitive computational cost. Barriers of key reaction steps must also be at or below the thermal energy available in the simulation to ensure sufficient sampling. Strategies that enhance sampling efficiency within AIMD include high-temperature acceleration[67], replica exchange[68], and metadynamics[69] to name a few. Alternatively, use of multiple, lower levels of theory can reduce computational cost and enable accelerated discovery, for example through the use of reactive force fields[70] or mixed quantum-mechanical/molecular mechanics techniques[71]. Recent work has also shown the possibility of discovering hundreds of new reaction pathways starting from a mixture of organic molecule species with AIMD sampling that is accelerated through high-temperature, relatively approximate first-principles methods, and variable boundary conditions.[72]

In order to elucidate the nature of early-stage reaction mechanisms in InP QD formation, we have employed variable boundary, high-temperature AIMD simulations starting from indium and phosphorus precursor mixtures. This approach enables us to observe the interactions and elementary reaction pathway steps that lead to formation of InP clusters. From these AIMD trajectories, we extract key reaction steps, and characterize their associated minimum energy pathways. Analysis of AIMD trajectories also reveals collective motions that contribute to the dynamic formation of an InP cluster. The outline of the paper is as follows. In section 2, we present the computational methods used in our study. In section 3, we present the results of our AIMD simulations including observation of the formation of a cluster, evaluation of intermediate structures, and determination of reaction pathway energetics based on AIMD trajectories. In section 4, we provide additional technical details into the acceleration strategies and approximations made in order to enable the sampling and observation of growth dynamics of InP QDs at reasonable computational cost. We provide our conclusions in section 5.

## 2. Computational Methods

AIMD simulations were performed with the graphical-processing unit (GPU)-accelerated quantum chemistry package, TeraChem[73-74]. Indium acetate (In(Ac)$_3$) and phosphine (PH$_3$) are used as the model indium and phosphorus precursors. A detailed discussion of the effect of precursor choice is presented in Sec. 4. One to seven In(Ac)$_3$ molecules and six to thirty PH$_3$ molecules are used in the AIMD simulations with the ratio between In(Ac)$_3$ and PH$_3$ molecules ranging from one to ten (Supporting Information Table S1). Twenty AIMD simulations were performed for a total simulation time of about 330 ps. Initial spherical AIMD configurations were generated using Packmol[75-76], with a minimum 3 Å van der Waals distance between individual molecules. AIMD simulations were carried out at constant temperature (T = 2000 K) using a Langevin thermostat with a damping time of 0.3 ps. The 5.5 to 40 ps Born-Oppenheimer MD simulations were carried out with a 0.5 fs time step. The electronic structure for AIMD simulations was evaluated at the Hartree-Fock (HF) level of theory with the 3-21G[77] basis set, and the impact of this method selection on the dynamics is discussed in Sec. 4.

Simulations were carried out with either constant or variable spherical boundary conditions. Molecules outside the initial radius, denoted $r_1$, experience a harmonic restraining force, while no force is applied to atoms inside the sphere. In the case of variable spherical boundary conditions, every 1.5 ps a Heaviside function was used to instantaneously decrease the boundary condition radius from $r_1$ to $r_2$ for 0.5 ps, after which the radius returned to its original value. The initial radius of the sampling space is chosen between 8 and 10 Å depending on the system size. For all variable spherical boundary condition simulations, the ratio of $r_2$ to $r_1$ was set to be between 0.6 and 0.7. More discussion of the effect of boundary condition choice is presented in Sec. 4. Coordination numbers evaluated for molecular dynamics trajectories are evaluated based on rescaled covalent radii (1.25x) of indium, phosphorus, oxygen and hydrogen atoms. These distance cut-offs are: 2.56 Å for In-O bonds, 3.16 Å for In-P bonds, and 1.79 Å for P-H bonds.[78] The values of these distance cutoffs also agree with the first local minimum on the radial distribution curve of In-O, In-P or P-H distance obtained from AIMD simulations.

Structures obtained from the HF AIMD trajectories were extracted for further evaluation with density functional theory (DFT) including geometry optimization and transition-state search and characterization. Since the synthesis of InP QDs is carried out using solvents with low dielectric constant ($\varepsilon$=2), all the energetic evaluations are performed in vacuum. DFT calculations in TeraChem employed the default B3LYP[79-81] functional, which uses the VWN1-RPA form[82] for the local density approximation component of the correlation. The composite basis set consisted of Los Alamos effective core potentials (LANL2DZ) for indium atoms and the 6-31G* basis set for all other atoms, which we refer to by the common short hand notation, LACVP*. Geometry optimizations of snapshots from AIMD trajectories were carried out in TeraChem using the DL-FIND[83] module with the L-BFGS algorithm in Cartesian coordinates. Default thresholds of $4.5 \times 10^{-4}$ hartree/bohr for the maximum gradient and $1.0 \times 10^{-6}$ hartree for change in self-consistent energy were employed. Partial charges were obtained from the TeraChem interface with the Natural Bond Orbital (NBO) v6.0 package[84]. NBO calculates the natural atomic orbitals (NAOs) for each atom by computing the orthogonal eigenorbitals of the atomic blocks in the density matrix, and the NBO partial charge on an atom is obtained as the difference between the atomic number and the total population for the NAO on the atom.

To characterize the energetics of different reactions, molecules that participated in reactive events were extracted from the overall AIMD trajectory and further processed to obtain transition state (TS) structures and activation energies. The reacting subset was identified based on observing changes in connectivity between molecules. Reactant and product structures were first geometry optimized using QChem 4.0[85] using the B3LYP/LACVP* level of theory. We validated our choice of functional and confirmed that alternative functionals did not substantially change predicted energetics (Supporting Information Table S2). Two energetic characterizations were carried out for bond rearrangement. In the case of isolated bond dissociation or formation events (e.g. indium precursor agglomeration, indium-phosphorus precursor adduct formation), constrained optimizations were carried out in which only the bond distance (e.g. In-O or In-P) was held fixed at values that span the relevant reaction coordinate. In the remaining cases (e.g. indium-precursor-assisted P-H bond dissociation), transition state searches were carried out in a multi-stage process. First, the freezing string method[86] was used to provide an interpolated path connecting the geometry-optimized reactant and product structures. In this method, linear synchronous transition (LST) is used for the initial interpolation of new path images that grow inward from reactant and product sides, and these structures are then optimized with the quasi-newton method with hessian update for up to a number of user specified steps (in our case, 3). The final number of images in the path approaches a target density (here, 21) but may exceed that value because a new image is guessed for the path until reactant- and product-derived paths cross. Unlike other path-based transition state search methods, the freezing string method does not obtain a converged minimum energy path but aims to provide a good guess for the location of a TS at reduced computational cost. The highest-energy image from the freezing string path was used as an initial guess for a partial rational function optimization (P-RFO)[87] TS search. These TSs were characterized with an imaginary frequency corresponding to the expected nuclear motion. For the dissociation processes involving multiple indium precursors, a second imaginary frequency with zero intensity and a value of <30i cm$^{-1}$ was obtained. Due to the abundance of soft modes in many of the structures, zero-point vibrational energy and entropy corrections were omitted. In cases where these effects could be included, activation energies changed by less than 2 kcal/mol. We also note that experimental solvents in QD synthesis consist of very low dielectrics. Inclusion of solvent effects[88-89] led to a lowering of activation energies by at most 2 kcal/mol. Tests for basis set superposition errors by computing the counterpoise correction[90] also led to less than 1 kcal/mol corrections on phosphine precursor energetics and were thus omitted.

## 3. Results and Discussion

### 3.1 Formation of an InP cluster during AIMD simulations

Using high-temperature ab initio molecular dynamics, we directly observe the formation of a small indium-rich InP cluster. Of twenty trajectories in total, this cluster formation is observed in an AIMD simulation that contains six In(Ac)$_3$ and six PH$_3$ molecules and constant boundary conditions (see Supporting Information Table S1). In this trajectory, overall cleavage of the three P-H bonds in a single PH$_3$ molecule results in the formation of a cluster with In$_4$P stoichiometry. While the cluster forms around only one PH$_3$ molecule, the other five PH$_3$ molecules rapidly adsorb and desorb from free indium sites. We estimate the maximum lifetime of adduct formation for each of the five PH$_3$ molecules as ranging from 0.08 to 0.60 ps at 2000 K, during which time the In-P distance is shorter than the previously defined distance cutoff. Four major interactions are observed during the overall cluster formation process: 1) agglomeration of the six In(Ac)$_3$ molecules into an [In(Ac)$_3$]$_6$ complex; 2) formation of an [In(Ac)$_3$]$_6$•PH$_3$ adduct; 3) dissociation of three P-H bonds concomitant with formation of In-P bonds; and 4) configurational rearrangement of the intermediates and cluster structures. For step 3, P-H bond dissociation is facilitated by nearby acetates of the indium precursors and leads to In-P bond formation. When the acetate that carries out hydrogen abstraction comes from the same precursor as that which is forming the In-P bond, we refer to this event as *intracomplex* P-H bond dissociation. Alternatively, the acetate that carries out hydrogen abstraction may belong to a different indium precursor, which we refer to as *intercomplex* P-H bond dissociation. Energetics for representative steps of each of these processes obtained at the B3LYP/LACVP* level of theory are discussed in detail in Sec. 3.2.

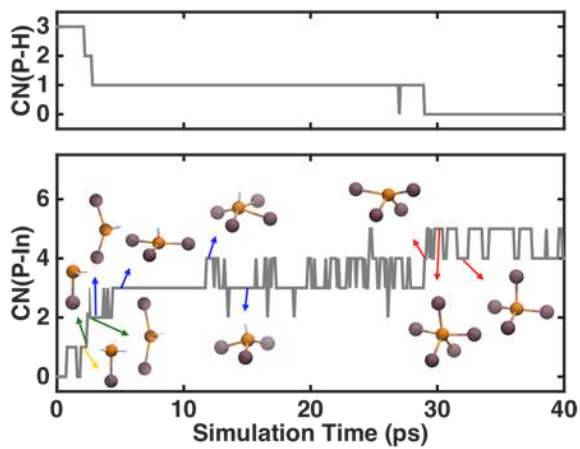

**Figure 2.** (top) Changes in coordination number of P by H and (bottom) P by In during cluster formation in an AIMD trajectory. Coordination numbers are plotted every 0.1 ps. Snapshots showing the phosphorus and the coordinating indium and hydrogen atoms are annotated in inset as indicated by arrows colored by the number of coordinating hydrogen atoms (three: yellow, two: green, one: blue, zero: red).

During the overall cluster formation process, the coordination environment of the reacting phosphorus precursor evolves from three hydrogen atoms to four or five indium atoms, as shown in Figure 2 (the evolution of In-P and P-H bond distances is available in Supporting Information Figure S1-2). In the first 4 ps of the high-temperature simulation, both the agglomeration and adduct formation processes occur, leading to significant rearrangement. During agglomeration, the six In(Ac)$_3$ molecules form a C-shaped chain in which individual precursors become linked by bridging carboxylates. Next, an [In(Ac)$_3$]$_6$•PH$_3$ adduct forms with the phosphorus atom weakly bound to an indium atom at one end of the chain, increasing the phosphorus coordination number (CN) from 3 to 4.

Next, dissociation of the first and second P-H bonds also occurs within the short 4 ps timeframe. After the adduct formation step, the first P-H bond is dissociated through the intercomplex mechanism, producing an HAc ligand and bringing the phosphorus CN back to 3 (Supporting Information Figure S3). This ligand remains bonded to the original indium precursor, but the abstracted proton rapidly transfers to a nearby acetate on the other end of the chain, rendering a new indium center undercoordinated. This second precursor then forms a second bond with the central phosphorus atom, producing a seesaw shaped complex with a central $In_2PH_2$ unit (H-P-H angle of 98$^o$ and In-P-In angle of 163$^o$). At the same time as the $In_2PH_2$ unit is formed, an acetate on one of the two phosphorus-coordinating indium sites changes from doubly coordinating indium (chelating bidentate) to single, monodentate coordination. This available oxygen anion is then free to abstract the second hydrogen atom from the central phosphorus (at around 2.7 ps) through the intracomplex pathway (Supporting Information Figure S4), leading to a three-coordinated phosphorus bonded with two indium atoms and one remaining hydrogen atom ($In_2PH$). In the remaining 1.3 ps, the central $In_2PH$ unit evolves from a nearly planar geometry to tetrahedral, with the lone pair on the P atom facing a nearby indium atom that then coordinates the central phosphorus, forming an $In_3PH$ unit at 4.0 ps. Such fast rearrangement suggests that these processes have barriers below $k_BT = 4.6$ kcal/mol at this level of theory.

After such rapid bond rearrangement, much slower rearrangement occurs over the next 25 ps, consisting primarily of slight rearrangement in the configuration of the central phosphorus unit and surrounding agglomerated indium precursors. During this interval, different phosphorus coordination types are observed including $In_2PH$, $In_3PH$, $In_4PH$, $In_5PH$ and $In_6PH$, with $In_3PH$ being the predominant configuration (around 72% of the time) followed by $In_4PH$ (around 25% of the time). These observations suggest a preference of phosphorus for CNs of 4 or 5. Along with the change of the phosphorus coordination, the two or three unbonded indium atoms gradually move from the side of the In$_n$PH unit to the top of the unit to form a cage-like structure that surrounds the remaining hydrogen atom (Supporting Information Figure S5). An available acetate must mediate the final P-H bond dissociation, but the agglomerated indium rearranges relatively slowly, causing the delay in the third P-H bond cleavage as compared to the first two P-H bond dissociations. We note, however, that the formation of the cage-like structure is likely only a necessity for phosphine precursors. If P(SiMe$_3$)$_3$ is instead used as the precursor, the P-Si bond distance is both longer and weaker, making it possible for more distant, intercomplex indium precursors to mediate bond dissociation.

This third P-H bond dissociation occurs through the intercomplex mechanism at around 28.9 ps (Supporting Information Figure S6). The formation of the fourth In-P bond occurs simultaneously with the P-H dissociation process, leading to the formation of a seesaw-geometry In$_4$P unit (Figure 2). Rearrangement of the cluster geometry is then observed after the dissociation of the third P-H bond until the end of the 40 ps AIMD simulation, with the phosphorus CN dynamically changing between four (~41% probability, approximate tetrahedral geometry) and five (~59% probability, approximate pyramidal or trigonal bipyramidal geometry). The average CN on the phosphorus atom is calculated to be around 4.6, slightly higher than that in the bulk, zinc blende InP crystals (CN = 4) but consistent with previous observations of coordination preference in other InP clusters[46].

We analyzed the In-P and In-O radial distribution functions from the last 10 ps of the AIMD trajectory (RDFs, see Supporting Information Figure S7). The In-P RDF has a first peak at 2.54 Å[91], identical to the experimental value of the In-P distance in bulk, zinc-blende InP crystals. Such good agreement confirms the fortuitous error cancellation observed between the use of HF and the near-minimal basis set with respect to higher-level theory (see Section 4 for more details). The In-O radial distribution function has a first peak at 2.12 Å, consistent with the In-O bond distance in an isolated B3LYP/LACVP*-optimized $In(Ac)_3$ molecule. We obtain a cluster structure by optimizing a snapshot taken at 35 ps in the AIMD simulation with B3LYP/LACVP* (Figure 3). This cluster structure has a central, tetrahedral core consisting of one phosphorus atom bonded to four indium atoms ($In_4P$). Importantly, the average In-P distance and In-P-In angle of the $In_4P$ unit match closely (within 0.01 Å and 1.2°, respectively) to experimental values for bulk, zinc-blende InP crystals. In addition to the core structure, two additional $In(Ac)_3$ molecules bond to other indium precursors through four shared, bridging bidentate acetates. The surface of larger InP QDs has been experimentally characterized as indium rich[18]. Our computational results are both consistent with this experimental observation but also demonstrate for the first time that indium rich clusters are formed even during the earliest stages of growth.

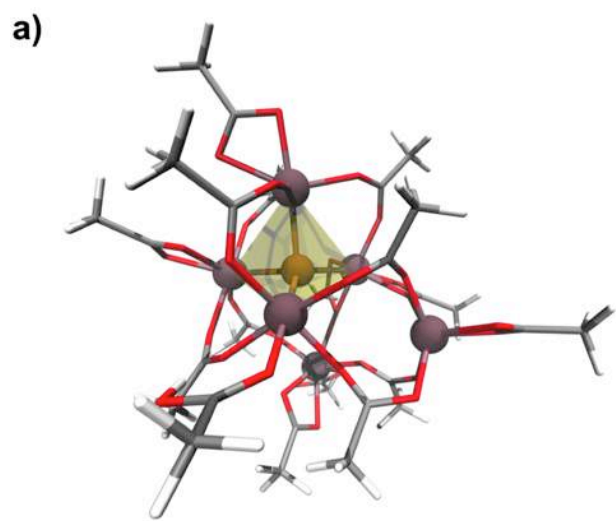

a)

b)

| $PIn_4$ | $d_{In-P}$ (Å) | $\theta_{In-P-In}$ (°) |
|---|---|---|
| Cluster | 2.55 | 108.3 |
| Crystal (exp.) | 2.54 | 109.5 |

**Figure 3.** (a) Structure of a cluster with tetrahedrally In-coordinated phosphorus obtained from geometry optimization of AIMD trajectory cluster (tetrahedron highlighted in yellow). (b) Comparison of average In-P bond distances and In-P-In angles in the cluster and experimental values for zinc blende, crystalline InP.

While the coordination environment around several of the indium atoms changes greatly over the course of the cluster formation process, shifts in B3LYP/LACVP* partial charges are modest (Supporting Information Figure S8). Instead, phosphorus and hydrogen atoms in the reacting phosphine molecule undergo significant changes of their electronic environment, with the partial charge of the phosphorus atom decreasing from 0.03 to -1.65 *e-* and the partial charge sum of the three hydrogen atoms increasing from -0.03 to 1.52 *e-*. These changes in partial charges occur alongside each P-H bond dissociation process. Relative charge values are also consistent between both B3LYP/LACVP* and HF/3-21G, in agreement with observations about error cancellation for precursors in this system (see Section 4). During the dissociation of the first P-H bond, the charge of the P atom drops by 0.52 *e-* while the charge of the leaving hydrogen atom increases by 0.51 *e-*. The partial charge of the indium atom bonded to the phosphorus atom slightly drops by about 0.09 *e-* (i.e., becomes more neutral) while partial charges of other indium atoms increase slightly (Supporting Information Table S3). During the dissociation of the second and third P-H bonds, a similar trend for the charge changes of phosphorus, hydrogen and indium atoms are observed (Supporting Information Table S4-5). These simultaneous and symmetric changes of phosphorus and hydrogen charges occur alongside the increase of the P-H bond distances, suggesting the polarization of the P-H bond during the dissociation process (Supporting Information Figure S9). Less drastic change of P charges would likely be observed during the dissociation of a P-Si bond when $P(SiMe_3)_3$ is used as the precursor because the P-Si bond is already more polar than P-H bond and P charges in P(SiMe$_3$)$_3$ (-0.71) are closer to the final value in the In$_4$P cluster (-1.65). Overall, this charge analysis suggests a partially ionic P-H dissociation mechanism mediated by the strongly negative charge localized on the abstracting carboxylate that causes evolution in the electronic structure around the phosphorus precursor but does not significantly affect the indium precursor.

### 3.2 Characterization of model reactions

Following analysis of the HF/3-21G AIMD trajectory that permitted direct observation of cluster formation, we now extract reaction steps sampled across trajectories and analyze energetics of indium agglomeration and In-P bond formation minimum energy pathways at the B3LYP/LACVP* level of theory.

#### a. Indium precursor agglomeration

Agglomeration of indium precursors (In(Ac)$_3$) to [In(Ac)$_3$]$_n$ complexes is observed in every AIMD trajectory with more than one indium precursor. During dynamics, complexes are formed through carboxylate ligands that bridge multiple indium atoms. We model the energetics for this agglomeration in the simplest case of two In(Ac)$_3$ molecules forming an [In(Ac)$_3$]$_2$ complex (Supporting Information Figure S10). In order to identify possible stable intermediates, we ran a 12 ps HF/3-21G AIMD simulation and geometry optimized 100 equally spaced snapshots in the last 10 ps with B3LYP/LACVP* (Supporting Information Table S6). The resulting intermediates are characterized by one to four bridging carboxylates shared between the two indium atoms. Five carboxylate ligand binding modes are observed: i) partial (3%) and ii) full (1%) monodentate (i.e. singly-coordinating indium) that are distinguished by the distance of the non-coordinating oxygen to the indium center; iii) chelating (52%), iv) bridging (31%), and v) chelating, bridging (13%) bidentate (i.e. doubly-coordinating indium), which are distinguished by the extent to which the oxygen atoms strongly coordinate the same indium atom (chelating) or are shared between two indium atoms (bridging) (Figure 4a). The observed binding modes are consistent with previous experimental and computational studies[18, 92], supporting indium agglomeration in the AIMD trajectories as mechanistically relevant.

When carboxylates bind in a monodentate fashion, the In-O bond is shortened (2.02 Å for partial mode and 1.93 Å for full mode) with respect to the three bidentate modes (2.11-2.26 Å), suggesting higher In-O bond order in monodentate cases. Monodentate modes are also thermodynamically unfavorable with respect to the bidentate binding modes, as suggested by their low occurrence frequencies. For example, the energetic cost of forming a full-monodentate acetate from a chelating-bidentate structure in the isolated precursor ($\Delta E_{FM-CB}$) is calculated as 21 kcal/mol (Supporting Information Figure S11). Notably, the carboxylate chain length has essentially no effect on the In-O bond energy, as $\Delta E_{FM-CB}$ for the longer chain indium myristate (In(My)$_3$, see Figure 1) is comparable at about 20 kcal/mol. The value of $\Delta E_{FM-CB}$ can be reduced in [In(Ac)$_3$]$_2$ complexes (as low as 15 kcal/mol) due to the stabilization offered from the bridging acetates (Supporting Information Figure S12). Further stabilization of monodentate carboxylates is likely in larger agglomerated complexes, as we previously noted several monodentate species that formed dynamically during the high-temperature AIMD simulations.

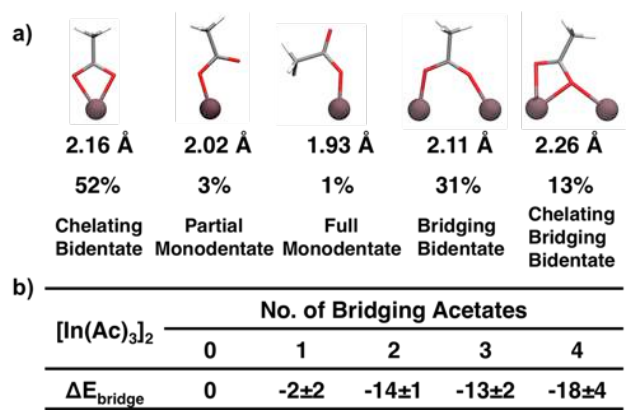

**Figure 4.** (a) Indium-carboxylate coordination modes obtained from optimized structures of MD-sampled [In(Ac)$_3$]$_2$ complexes annotated with average In-O distance and frequency of occurrence. (b) Relative energies of chelating bridging bidentate or bridging bidentate [In(Ac)$_3$]$_2$ complex structures with increasing numbers of bridging bidentate ligands.

As the number of bridging ligands between In centers increases, the formation energy of the [In(Ac)$_3$]$_2$ complex becomes increasingly favorable. While a single bridging acetate ligand is only favorable by -2 ± 2 kcal/mol, the energetic benefit increases to -18 ± 4 kcal/mol for four bridging acetates (Figure 4b). Ranges in energetics here are obtained from the small variations in 100 geometry-optimized snapshots. We have also computed energetic barriers for the sequential formation of bridging ligands. The initial agglomeration of two indium precursors to form a complex with one to three bridging acetates has a low activation energy (0-3 kcal/mol), while the barrier for adding a fourth bridging acetate to the complex is calculated to be 6 kcal/mol (Supporting Information Figure S13). The higher barrier due to formation of the fourth bridging interaction is largely due to the fact that the three-bridge structure is more stable than the four-bridge structure by about 2 kcal/mol (i.e. the reverse barrier for transitioning from four- to three-bridge structures is only 4 kcal/mol). In each step, the new acetate bridge forms by orienting toward the neighboring indium atom in a position compatible with bonding followed by shortening of the second indium-oxygen bond.

Overall, observations of both energetically favorable and low-barrier rearrangement for agglomeration through bridging interactions supports our previous observations of dynamic indium agglomeration during cluster formation. We note, however, that indium precursors employed during experimental synthesis (e.g., indium myristate) may be less likely to form the highly-interconnected four-bridge structures due to steric repulsion of the long alkyl chains. Importantly, the observation of the reduction in energetic penalty for monodentate indium-carboxylate ligands with increasing agglomeration is likely preserved even in the case of long chains. This result suggests that agglomeration of the indium precursor may facilitate creation of indium sites available for forming In-P bonds.

### b. Formation of In-P Bonds

In addition to indium precursor agglomeration, AIMD sampling of mixtures of $In(Ac)_3$ and $PH_3$ precursors reveals multiple modes of In-P bond formation. In all cases, $PH_3$ molecules first weakly associate with individual $In(Ac)_3$ or complexed $[In(Ac)_3]_n$ precursors to form $In(Ac)_3 \bullet PH_3$ or $[In(Ac)_3]_n \bullet PH_3$ adducts. We previously noted that this weak interaction was characterized during dynamics by relatively short, sub-ps lifetimes. For stable In-P bond formation to occur, P-H bond dissociation must occur. When the In-P bond formation and P-H bond cleavage is mediated by a carboxylate coordinated to the indium forming the In-P bond, this process is called *intracomplex*, while we refer to the process as *intercomplex* when the acetate is coordinated to a different indium precursor (Scheme 1). We now examine possible mechanistically relevant differences in the energetics between the two types of P-H bond dissociation pathways.

**Scheme 1**. (a) Intracomplex P-H bond dissociation mechanism with In(Ac)$_3$ or [In(Ac)$_3$]$_2$ both participating in the In-P bond formation and carrying out hydrogen abstraction. (b) Intercomplex P-H bond dissociation mechanism in which a second In(Ac)$_3$ carries out the hydrogen abstraction.

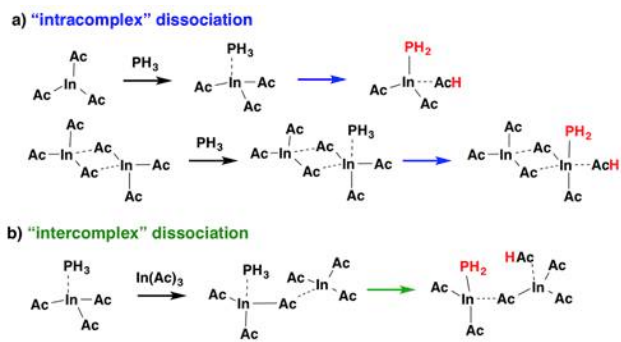

*Intracomplex pathway*. We first consider the formation of an In(Ac)$_3$•PH$_3$ adduct and subsequent intracomplex P-H bond dissociation (Scheme 1a, Supporting Information Figure S14, and Figure 5). The optimized In(Ac)$_3$•PH$_3$ adduct has an In-P distance of 3.69 Å, confirming the weak interaction between In and P observed in AIMD trajectories. Following the barrierless adduct formation, one of the In-O bonds breaks (d$_{In-O}$=3.65 Å) and the uncoordinated oxygen abstracts a proton from PH$_3$ to form an In-P bond (d$_{In-P}$=2.61 Å). The In-O and In-P bond distances closely resemble their values in the product of 3.77 Å and 2.55 Å respectively, which would initially suggest the formation of a late transition state. However, the transferring proton is shared between phosphorus and oxygen (d$_{P-H}$=1.62 Å, d$_{O-H}$=1.34 Å) in the transition state (Supporting Information Table S7). The activation energy for the In-P bond formation process is 21 kcal/mol, which is the same as the energy penalty to form a full-monodentate acetate from a chelating-bidentate acetate in In(Ac)$_3$.

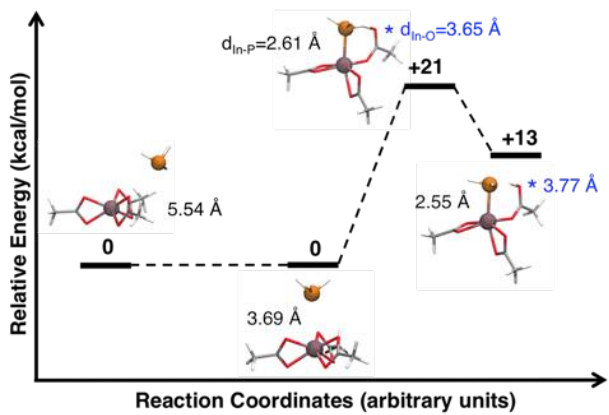

**Figure 5.** Energetics and structures of the intracomplex pathway with one In(Ac)$_3$ and one PH$_3$ molecule for In-P bond formation. Indium-phosphorus distances are shown in black and indium-oxygen distances are shown in blue for the unbonded oxygen that is denoted with a blue asterisk.

Since the energetic cost of forming monodentate acetates is lowered in agglomerated indium precursors, we computed the energetics of an intracomplex pathway using a two-bridge [In(Ac)$_3$]$_2$ complex (Scheme 1a, Supporting Information Figure S15 and Figure 6). Here, we observe an adduct with a shorter In-P bond (2.85 Å) that is mediated by partial lengthening of one of the indium-acetate oxygen bonds (3.01 Å). The process of forming the [In(Ac)$_3$]$_2$•PH$_3$ adduct destabilizes the uncoordinated oxygen, requiring 5 kcal/mol. After the adduct formation, the destabilized oxygen further dissociates from indium (d$_{In-O}$=3.68 Å) in order to abstract a proton from the associated phosphine, with a barrier of 15 kcal/mol. The overall barriers for the process are 20 kcal/mol, yielding no net energetic benefit of agglomeration on the activation energy for In-P bond formation in this case. However, we did observe during dynamics that agglomeration of larger numbers of indium precursor copies increased the likelihood of dynamic formation of monodentate acetates. Further analysis in this case reveals that although monodentate mode is more stabilized in the [In(Ac)$_3$]$_2$ complex, the proton affinity of the uncoordinated oxygen in the [In(Ac)$_3$]$_2$ complex is lowered as compared to In(Ac)$_3$ (Supporting

Information Table S8). The cancellation of these two effects produces comparable activation energies and transition state structures for the two intracomplex dissociation processes. Notably, in both cases, In-O and In-P bond distances more closely resemble the products than the adduct distances, but the transferring hydrogen is located symmetrically between P and O (Supporting Information Table S7).

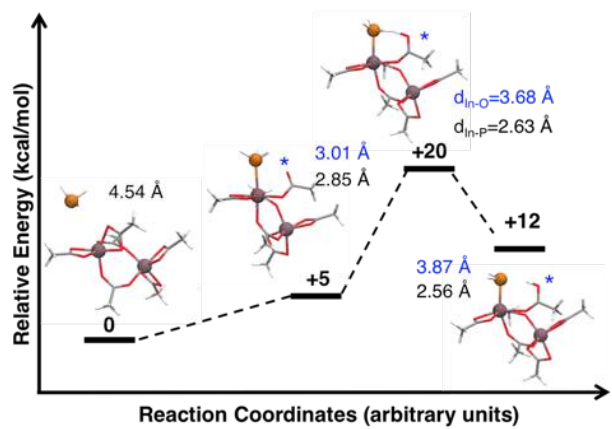

**Figure 6.** Energetics and structures of the intracomplex pathway between a $[In(Ac)_3]_2$ complex and one $PH_3$ molecule for In-P bond formation. Indium-phosphorus distances are shown in black and indium-oxygen distances are shown in blue for the unbonded oxygen that is denoted with a blue asterisk.

*Intercomplex pathway*. We modeled the intercomplex pathway using one $In(Ac)_3$•$PH_3$ adduct and a separated $In(Ac)_3$ precursor for comparison with the intracomplex pathway (Scheme 1b, Figure 7, and Supporting Information Figure S16). The formation of a bridging acetate between the adduct and $In(Ac)_3$ has a reaction barrier of 4 kcal/mol due to the partial dissociation of an In-O bond to form a bridging-bidentate acetate. The subsequent P-H bond dissociation is facilitated by the cleavage of another In-O bond in the nearby $In(Ac)_3$, after which the uncoordinated oxygen abstracts a hydrogen atom from $PH_3$, leading to the formation of an In-P bond. The transition state is characterized by a 0.5 Å shorter distance between indium and the under-coordinated oxygen ($d_{In-O}$=3.2 Å) than that in the corresponding transition state from the intracomplex pathway, although P-H and O-H distances are comparable (Supporting Information Table S7). The shorter In-O bond distance lowers the activation energy by 7 kcal/mol with respect to the intracomplex pathway. This relationship between distance and relative energetics agrees well with the evaluated In-O bond distance dependence of energetics in isolated In(Ac)$_3$ molecule where a configuration with a 3.2 Å In-O bond is 5 kcal/mol lower in energy than one with a 3.7 Å In-O bond (Supporting Information Figure S11). The favorable geometry afforded in this intercomplex pathway calculation implicates the dominance of intercomplex pathways over the intracomplex pathway, consistent with AIMD trajectories in which the majority of In-P bond formation steps occurred through the intercomplex pathway. This observation also highlights the important role of bridging coordination modes to both stabilize indium while predisposing an oxygen atom toward proton abstraction from the phosphorus precursor.

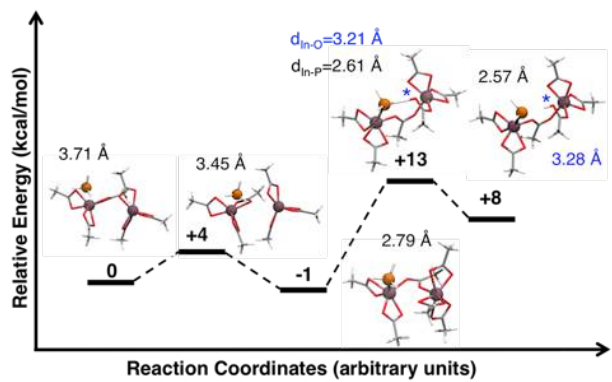

**Figure 7.** Energetics and structures of the intercomplex pathway for In-P bond formation. Indium-phosphorus distances are shown in black and indium-oxygen distances are shown in blue for the unbonded oxygen that is denoted with a blue asterisk.

We have noted that In-P bond formation and P-H dissociation pathways have energy barriers corresponding roughly to the energetic cost of forming a monodentate acetate on an indium precursor. Therefore, starting from an indium precursor structure with pre-formed monodentate acetates should lower the energy barrier for the formation of In-P bonds. We designed a new intracomplex pathway starting with a full-monodentate In(Ac)$_3$ and PH$_3$ molecule (Supporting Information Figure S17-18) and confirmed that the barrier of new pathway is calculated to be 5 kcal/mol. Although the reaction barrier is lowered, we have simply shifted the cost to form a full-monodentate In(Ac)$_3$ away from the P-H dissociation transition state and have not reduced the overall steepness of the reaction landscape. However, dynamic formation of monodentate acetates is likely under experimental conditions, particularly if ligands are designed that stabilize the process of forming undercoordinated indium species by shifting the relative energetics of bidentate and monodentate binding modes.

Since the experimental phosphorus precursor (P(SiMe$_3$)$_3$) has different geometric and electronic properties than PH$_3$, we expect that these differences should affect In-P bond formation energetics (Supporting Information Figure S19-20). Here, the adduct formation is again barrierless. However, the In-P distance (2.75 Å) is 0.94 Å shorter in this adduct than in In(Ac)$_3$•PH$_3$ due to the additional hydrogen bonding interactions between methyl groups of P(SiMe$_3$)$_3$ and indium precursor acetate ligands (Supporting Information Figure S21). As a result, the formation of the In-P bond not only requires the cleavage of an In-O bond but also disrupts some of the favorable hydrogen bonding interactions. The loss of these hydrogen bonding interactions explains why a higher estimate of the activation energy (27 kcal/mol) is observed for the larger precursor, despite the lower bond dissociation energy of the P-Si bond. We have not computed the energetics of the intercomplex pathway due to computational cost and large number of soft degrees of freedom, but we expect that the reaction energy will be significantly lowered both due to the lower activation energy for that pathway with phosphine and due to the preservation of favorable hydrogen bonding interactions. While In-P bond formation with phosphine precursors was endothermic by 13 kcal/mol, it becomes exothermic by 5 kcal/mol for P(SiMe$_3$)$_3$ precursors, consistent with differences in the bond dissociation energy of the P-H and P-Si bonds in the two molecules. A clear mechanistic picture emerges that the phosphorus-ligand bond in the phosphorus precursor controls reaction thermodynamics for In-P bond formation, while the indium-ligand bond controls the activation energy, and thus kinetics, for the process. Thus, while experimental efforts to tune InP QD synthesis have focused on phosphorus precursor chemistry[22] without significant effect on quality of QDs, a concerted effort that tunes both indium and phosphorus precursors is likely necessary.

**4. Acceleration approach**

In order to accelerate the fruitful sampling and discovery of pathways that lead to formation of InP QDs, we have made a number of careful approximations. We now consider what effect such approximations may have on the predicted mechanism of cluster growth by considering a) the effect of model precursor choice, b) the role of the electronic structure method and basis set size in AIMD, and c) the choice of boundary conditions employed.

**a. Choice of model precursors**

Long chain indium carboxylates such as indium myristate (In(My)$_3$, 130 atoms) and tris(trimethylsilyl) phosphine (P(SiMe$_3$)$_3$, 40 atoms) are the most commonly used precursors for the synthesis of InP QDs[19-21, 24-26, 93] (see Figure 1). Due to the large size of these experimental precursors, we employ indium acetate (In(Ac)$_3$, 22 atoms) and phosphine (PH$_3$, 4 atoms) as model molecules (see Figure 1). Our choice is justified in part by the fact that phosphine has been used as a precursor with In(My)$_3$ to grow InP QDs[94], and previous computational QD studies have often employed acetates to model longer chain carboxylates[41, 92]. The speed-up obtained by using the model precursors is substantial: HF/3-21G gradient calculations on In(Ac)$_3$ are about 6 times faster than In(My)$_3$ and on PH$_3$ are about 100 times faster than P(SiMe$_3$)$_3$.

In order to identify whether the simplifications of the model precursors affect the electronic and geometric structure, we compared properties of the simplified molecules with their experimental analogues. For the two indium precursors, identical In or O partial charges and In-O bond distances are obtained regardless of the use of HF/3-21G or B3LYP/LACVP* as the electronic structure method (Supporting Information Table S9). We investigated whether the steric effect of longer chain carboxylates alters dynamics by carrying out AIMD simulations with bulkier indium trimethyl acetate (In($^t$BuCOO)$_3$, 49 atoms) and PH$_3$ precursors. Comparable reaction pathways were observed in AIMD trajectories of In($^t$BuCOO)$_3$ and In(Ac)$_3$ (Supporting Information Figure S22).

Unlike the indium precursors, the B3LYP/LACVP* partial charges of phosphorus in PH$_3$ and P(SiMe$_3$)$_3$ substantially differ at 0.01 *e*- and -0.71 *e*-, respectively. This difference is expected since P and H atoms have comparable electronegativity[95], giving rise to covalent bonding, while the less electronegative Si leads to a partially ionic P-Si bond. The P-H bond is also 0.86 Å shorter than the P-Si bond, suggesting a stronger interaction between P and H atoms as compared to that between P and the SiMe$_3$ group that is also reflected in the B3LYP/LACVP* bond dissociation energies: 78 kcal/mol for the P-H bond and 62 kcal/mol for the P-Si bond. Nevertheless, AIMD simulations of In(Ac)$_3$ and P(SiMe$_3$)$_3$ molecules generated analogous intermediates and reaction steps as when PH$_3$ was used (Supporting Information Figure S22).

Reaction coordinate characterization in this work focuses on $PH_3$ precursors, but both predicted and calculated differences with respect to $P(SiMe_3)_3$ precursors will be discussed where relevant.

**b. Accelerated ab initio molecular dynamics sampling**

We have employed a number of strategies to enhance the rate of sampling of reactive collisions between indium and phosphorus precursors. In addition to employing an elevated temperature of 2000 K, we carry out dynamics with a low level of theory and near-minimal basis set combination (i.e., HF/3-21G). Snapshots from the dynamics are then refined by B3LYP/LACVP* geometry optimizations and minimum energy path searches. Therefore, the HF/3-21G AIMD sampling only needs to reproduce qualitative changes in bonding consistent with B3LYP/LACVP*. The computational benefit of using HF/3-21G is significant: for an example 156 atom simulation that contains 6 indium precursors carried out in TeraChem on 2 GPUs, HF/3-21G requires about 60 s per MD step while B3LYP/LACVP* requires about 1900 s per MD step. We note that the predominant computational cost in all simulations presented in this work is associated with the electron-rich indium precursors.

It is useful to quantify, however, the extent to which HF, which includes no treatment of dynamic correlation, produces a substantially different description of the electronic structure than a hybrid DFT approach. First, we compare the equilibrium properties of $In(Ac)_3$ and $PH_3$ precursors (Figure 8). HF yields shorter In-O and P-H bond distances than B3LYP at the same basis set size by about 0.05 and 0.03 Å, respectively. However, both HF and B3LYP bond distances are reduced when increasing the basis set size from 3-21G to LACVP* by about 0.03 to 0.05 Å, with B3LYP reductions slightly larger than HF. The net result is a cancellation of errors between the underestimation of bond lengths by HF and overestimation from the incomplete basis, leading to agreement within 0.01 Å between HF/3-21G and B3LYP/LACVP* for In-O and P-H bond lengths. Interestingly, partial charges computed with NBO show a similar trend. For the same basis set, HF overpolarizes the In-O bond compared to B3LYP. For a given method, increasing the basis set size leads to larger In-O charge separation, and B3LYP is more sensitive to the increase in basis set size than HF is. The net result is that while the In partial charge with HF/LACVP* of around +2.2 $e$- overestimates the B3LYP/LACVP* In charge by +0.2 $e$-, HF/3-21G and B3LYP/LACVP* partial charges are in quantitative agreement. Similar fortuitous cancellation of errors between basis set incompleteness and absence of treatment of dynamic correlation is apparent in the P atom charges in $PH_3$. Agreement of vibrational frequencies is also good, as comparison of frequencies for the optimized $In(Ac)_3$ molecule using HF/321G and B3LYP/LACVP* reveals that the absolute mean difference for frequencies larger than 50 cm$^{-1}$ is around 9% (Supporting Information Table S10). This overall fortuitous error cancellation therefore further motivates sampling AIMD at the HF/3-21G level at around 1/30$^{th}$ of the computational cost of a B3LYP/LACVP* calculation.

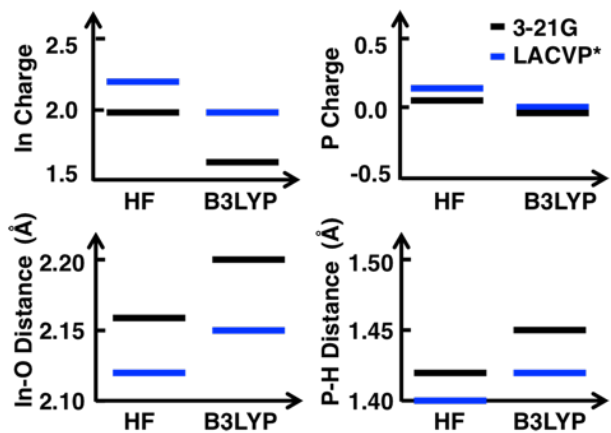

**Figure 8.** Comparison of partial charges (In or P, top) and bond distances (bottom) for In(Ac)$_3$ (left) and PH$_3$ (right) molecules obtained with Hartree-Fock or DFT (B3LYP) using the 3-21G (black line) or LACVP* (blue line) basis sets.

While equilibrium properties are consistent between HF/3-21G and B3LYP/LACVP*, our ultimate interest is in reproducing dynamics and sampling of reactive intermediates, which necessitates moving away from equilibrium, zero-temperature properties. Parallel AIMD simulations with both HF/3-21G and B3LYP/LACVP* were performed for a system containing one In(Ac)$_3$ and one PH$_3$ molecule. Nearly identical In-O and P-H radial distribution function peak values and shapes were obtained from the two simulations (Supporting Information Figure S23). As an even more strenuous test case, at transition-states, basis set superposition errors and static correlation errors can dominate and interfere with the error cancellation. For the formation of an In-P bond from an In(Ac)$_3$•PH$_3$ adduct to In(Ac)$_2$PH$_2$•HAc, HF/3-21G predicts both a lower barrier (~6 kcal/mol less) and more favorable reaction energy (~13 kcal/mol lower). This discrepancy in barrier heights is not necessarily a disadvantage, as lower barriers for desired reactions will lead to enhanced sampling of reaction events during the AIMD simulation. Regardless, all AIMD trajectories obtained with HF/3-21G are then quantitatively assessed with B3LYP/LACVP* through geometry optimization and transition state searches.

c. **The choice of boundary conditions**

For reaction mechanism discovery involving organic species, it has recently been shown[72] that enforcing spherical boundary conditions that shrink periodically is a useful way to enhance the frequency of collisions during molecular dynamics that lead to chemical transformations. As the chemistry of the systems studied here are quite distinct from small organic molecules, we have tested the extent to which periodically shrinking boundary conditions is a useful strategy to enhance the sampling rate of reactive collisions between indium and phosphorus precursors. In this approach, the key adjustable parameters in this approach are the ratio ($r_2/r_1$) of the smaller radius ($r_2$) to the larger radius ($r_1$), the time spent at each radius ($t_1$ versus $t_2$), and the harmonic restraint force applied ($k_1$ versus $k_2$) to enforce the boundary conditions. We have chosen $r_2/r_1$ to be around 0.6-0.7 since a tighter $r_2$ can lead to the unphysical cleavage of C-C and C-H bonds. We note that the shrinking process also slows convergence to self-consistency and increases the time per MD step. For example, in a simulation containing 6 indium precursors and 6 phosphorus precursors, the time per MD step increases by a factor of 1.7 when the smaller boundary is enforced. For the other two parameters, we employ the defaults outlined in Ref. [72], which correspond to $t_1$=1.5 ps and $t_2$=0.5 ps and force constants $k_1$=1.0 kcal/(mol•Å$^2$) and $k_2$=0.5 kcal/(mol•Å$^2$). One further disadvantage of this approach is the loss of direct information about the timescale of events, but since we are already using a lower level of theory (HF/3-21G), direct dynamics timescales already carry limited meaning.

Generally, we have observed different growth intermediates when using different boundary conditions. For example, in two AIMD simulations containing one $In(Ac)_3$ and eight $PH_3$ molecules (Supporting Information Figure S24), the use of shrinking boundary conditions leads to the formation of an intermediate species $In(Ac)_2PH_2$ while the use of constant boundary conditions only leads to the formation of an $In(Ac)_3•PH_3$ adduct. Conversely, in another two AIMD simulations sampling the interaction between three $In(Ac)_3$ and thirty $PH_3$ precursors, only the intermediate structures such as $[In(Ac)_3]_3$ complexes and $[In(Ac)_3]_3•PH_3$ adducts were observed in a 15 ps simulation at periodically shrinking boundary conditions without explicit In-P bond formation. Employing constant boundary conditions for the same system leads to the formation of various different intermediates containing stable In-P bonds within 12 ps (Figure 9).

Although [In(Ac)$_3$]$_3$•PH$_3$ is still observed, further interaction between an acetate and P-H bond within the [In(Ac)$_3$]$_3$•PH$_3$ adduct leads to the first P-H bond dissociation at about 5 ps followed by a second P-H bond dissociation at about 8 ps. Mechanisms and dynamics of In-P bond formation are discussed in detail in the main body of the text. However, these observations highlight the fact that for the more complex inorganic systems we are interested in, an effective sampling strategy must incorporate varying absolute and relative numbers of indium and phosphorus precursors and applying both variable and constant boundary conditions.

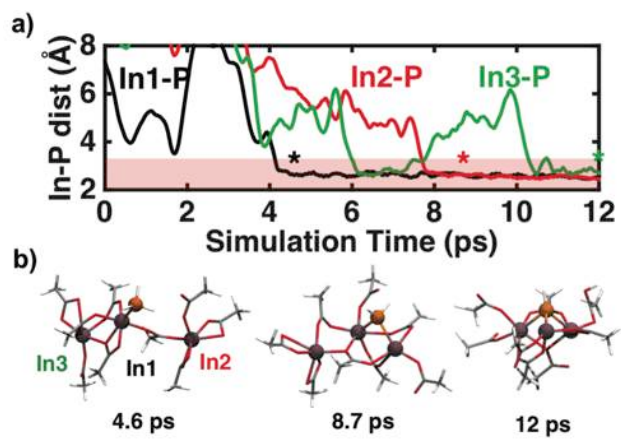

**Figure 9.** (a) Evolution of In-P distances in a constant spherical boundary AIMD trajectory for a single PH$_3$ molecule that forms In-P bonds with three In(Ac)$_3$ molecules. The distance cutoff for In-P bonding is indicated as a shaded region in the graph. (b) Snapshots of only the reacting molecules from the trajectory at different simulation times as annotated by color-coded asterisk in (a).

## 4. Conclusions

We have presented a computational approach for the sampling and discovery of reactive intermediates that form during early stage growth of indium phosphide quantum dots. As the structure of intermediates were previously unknown, we undertook a number of efforts to ensure sampling of possible favorable configurations over a total of 330 ps of AIMD on systems up to 277 atoms in size. This acceleration strategy included the use of GPU-accelerated quantum chemistry employed without explicit dynamic correlation (i.e. Hartree-Fock) and in a near-minimal basis set (3-21G). Opposing effects of inclusion of dynamic correlation (i.e. with the hybrid functional B3LYP) and a larger LACVP* polarized basis set led to fortuitous error cancellation that imparted excellent agreement between HF/3-21G and B3LYP/LACVP* geometric and electronic structure, allowing us to directly sample dynamics at $1/30^{th}$ of the computational cost of a production-quality DFT calculation. Additionally, we employed variable boundary conditions in order to enhance diversity and number of reactive collisions that were sampled in our high-temperature molecular dynamics.

Our sampling strategy enabled us to directly observe the formation of an indium-rich $In_4P$ cluster that already possesses the same structural properties as the experimental bulk zinc-blende crystal structure of indium phosphide. Our indium rich cluster is consistent with experimental characterization of larger InP QDs, and we demonstrate for the first time that an indium-rich surface is likely present from the earliest stages of growth. During the 40 ps cluster formation trajectory, we observed rapid agglomeration of indium precursors around a single phosphorus precursor. In these simulations, cluster formation was mediated by cooperative effects, which we refer to as an intercomplex pathway, in which one indium precursor formed a bond with the phosphorus center while another abstracted a proton from phosphine. We then characterized the minimum energy pathways of key processes observed in dynamics and confirmed that intercomplex pathways for In-P bond formation was also found to be more energetically favorable than the intracomplex pathway due to stabilization of the indium-acetate oxygen during the proton abstraction process.

Overall, we consistently observed that the highest barriers to In-P bond formation were exclusively determined by energetic penalties associated with indium-carboxylate bond cleavage. The net favorability of the reaction, on the other hand, was determined by the nature of the phosphorus precursor (e.g. $PH_3$ or $P(SiMe_3)_3$). These observations challenge the paradigm that a single precursor may be tuned in order to optimize the target size distribution of InP QDs. Altering the stability of agglomerated and monodentate indium precursor structures will adjust the energy landscape for In-P bond formation, while tuning phosphorus precursor chemistry can alter how much heat is generated during the QD synthesis process by the downhill nature of the precursor reaction. In the future, a greater focus on tuning the chemistry of indium precursors should enhance the quality of InP QDs with desired size distributions and enable the growth of the use of InP QDs in a wide range of consumer applications.

## ASSOCIATED CONTENT

**Supporting Information Available**: Summary of the AIMD parameters and example intermediates; Comparison of NBO charges, optimized geometries, frequencies, and dynamic properties of indium precursors evaluated using HF and B3LYP; Summary of the geometries and energies of the one hundred $[In(Ac)_3]_2$ structures; Energy evaluation of the agglomeration of indium precursors, the formation of adducts and the dissociation pathways of P-H/P-Si bonds; Geometries and NBO charge changes during the cluster formation process. This material is available free of charge via the Internet at http://pubs.acs.org.

## AUTHOR INFORMATION

**Corresponding Author**

*email:hjkulik@mit.edu  phone:617-253-4584

**Notes**

The authors declare no competing financial interest.

## ACKNOWLEDGMENT

This work was supported by the National Science Foundation under grant number ECCS-1449291. H.J.K. holds a Career Award at the Scientific Interface from the Burroughs Welcome Fund. This work was carried out in part using computational resources from the Extreme Science and Engineering Discovery Environment (XSEDE), which is supported by National Science Foundation grant number ACI-1053575. This work was also carried out in part using the computational resources of the Center for Nanoscale Materials (Carbon cluster), an Office of Science user facility, was supported by the U. S. Department of Energy, Office of Science, Office of Basic Energy Sciences, under Contract No. DE-AC02-06CH11357.

**Table of Contents**

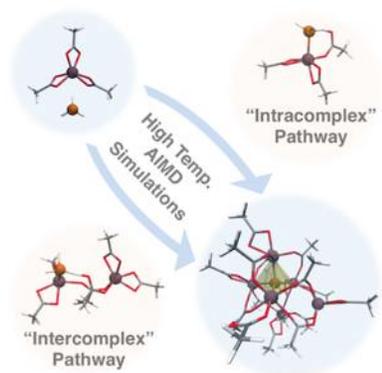